\documentclass[aps,prb,twocolumn]{revtex4}
\usepackage{graphicx}
\usepackage{bm}
\usepackage{wasysym}

\begin{document}

\title{Three-dimensional topological phases in a layered honeycomb spin-orbital model}
\author{Gia-Wei Chern}
\affiliation{Department of Physics, University of Wisconsin,
Madison, Wisconsin 53706, USA}

\begin{abstract}
We present an exactly solvable spin-orbital model based on the
Gamma-matrix generalization of a Kitaev-type Hamiltonian. In the
presence of small magnetic fields, the model exhibits a critical
phase with a spectrum characterized by topologically protected Fermi
points. Upon increasing the magnetic field, Fermi points carrying
opposite topological charges move toward each other and annihilate
at a critical field, signaling a phase transition into a gapped
phase with trivial topology in three dimensions. On the other hand,
by subjecting the system to a staggered magnetic field, an effective
time-reversal symmetry essential to the existence of
three-dimensional topological insulators is restored in the
auxiliary free fermion problem. The nontrivial topology of the
gapped ground state is characterized by an integer winding number
and manifests itself through the appearance of gapless Majorana
fermions confined to the two-dimensional surface of a finite system.
\end{abstract}

\maketitle

\section{Introduction}

Topological phases of matter are one of the most remarkable
discoveries in modern condensed-matter physics.
\cite{tknn,avron83,kohmoto85,wen90,volovik} Instead of a local order
parameter describing broken symmetries of the system, this novel
state of matter is characterized by a topological invariant which is
insensitive to small adiabatic deformations of the Hamiltonian. A
classic example is the integer quantum Hall (IQH) effect, where the
quantized Hall conductance corresponds to a topological invariant
called the first Chern number or the TKNN integer. \cite{tknn}
Another distinctive property of IQH states is the appearance of
zero-energy modes at the sample edge, despite the fact that all bulk
excitations are fully gapped. Due to their topological nature, these
conducting edge modes persist even in the presence of disorders.
Recently, theoretical investigations have shown that similar
topological insulators can be generalized to time-reversal invariant
systems (quantum spin-Hall effect)
\cite{kane05,bernevig06,moore07,roy09a} and to three dimensions.
\cite{fu07a,fu07b,roy09b} Signatures of topological insulators such
as quantized conductance and protected surface Dirac cone have been
reported experimentally in semiconducting alloys and quantum wells.
\cite{konig07,hsieh08,xia09,roushan09,chen09} Following these
developments, systematic classifications of topological insulators
have also been proposed. \cite{schnyder08,kitaev09,qi08}

Although most discussions of topological insulators are in the
context of tight-binding fermionic models or mean-field
superconductors, it has been shown that topological insulators can
also emerge from strongly correlated electronic systems.
\cite{jackeli09,shitade09,raghu08,sun09,zhang09} An exactly solvable
example is Kitaev's anisotropic spin-1/2 model on the
two-dimensional honeycomb lattice. \cite{kitaev06} As shown in his
seminal paper, the spin model can be reduced to a problem of free
Majorana fermions coupled to a static $Z_2$ gauge field. The ground
state of the Kitaev model has two distinct phases. The gapped
Abelian phase is equivalent to the toric code model \cite{kitaev03}
whose excitations are Abelian anyons. Relevant to our discussion is
the non-Abelian B phase in the presence of a magnetic field. This
phase is characterized by an integer winding number $\nu = \pm 1$.
Similar to IQH insulators, the non-Abelian phase also supports
gapless chiral edge modes (whose chirality depends on the sign of
magnetic fields) except that the edge modes here are real Majorana
fermions, as contrasted to complex fermions in the case of IQH
states.

There has been much effort to generalize Kitaev model to other
trivalent lattices \cite{yang07,yao07} and to three-dimensions.
\cite{si07,si08,mandal09} Probably the most notable example is the
discovery of a chiral spin liquid as the ground state of Kitaev
model on a decorated-honeycomb lattice. \cite{yao07} On the other
hand, despite being exactly solvable, we find that most gapped
phases of 3D Kitaev model is topologically trivial. \cite{chern09}
The fact that the model can only be defined on lattices with
coordination number 3 significantly constrains the possible Majorana
hopping Hamiltonian. Recently, noting that the exact solvability of
Kitaev model relies on the fact that the three spin-1/2 Pauli
matrices realize the simplest (dimension-2) Clifford algebra, the
so-called $\Gamma$-matrix generalization of the Kitaev model offers
richer possibilities of engineering exactly solvable models with
unusual phases in both 2D and 3D.
\cite{levin03,hamma05,yao09,wu09,ryu09,nussinov09} Physically,
models based on, e.g. dimension-4 $\Gamma$ matrices can be
interpreted as spin-$\frac{3}{2}$ models,
spin-$\frac{1}{2}\frac{1}{2}$ models, or spin-orbital models.

Based on the above $\Gamma$-matrix generalization of  Kitaev model,
emergent topological insulators have been demonstrated on a 3D
diamond lattice with coordination number 4. \cite{wu09,ryu09} To
construct a Kitaev-type model on such a lattice, one needs four
mutually anticommuting operators for the four nearest-neighbor
links. This can be easily realized using the dimension-4
$\Gamma$-matrix representation of the Clifford algebra. By
exploiting the redundancy of representing two such sets of bosonic
operators in terms of 6 Majorana fermions, the diamond-lattice model
enjoys an effective time-reversal symmetry (TRS) which is essential
to the existence of topological insulators in 3D. \cite{schnyder08}
With only nearest-neighbor interactions, the diamond-lattice model
reduces to a problem of two identical copies of free Majorana
fermions sharing the same $Z_2$ gauge field. The permutation
symmetry between the two fermion species thus manifests as a TRS. In
the weak-pairing regime of the model, hybridization of the two
Majorana species results in a gapped ground state with nontrivial
topology.

A natural question to ask is whether it is possible to realize
the required TRS in a Kitaev-type model via generic physical
mechanisms. In this paper, we provide such an example through a
natural generalization of the honeycomb Kitaev model. Using a
similar $\Gamma$-matrix formalism, we study a Kugel-Khomskii-type
\cite{kk} spin-orbital Hamiltonian on a layered honeycomb lattice
whose coordination number is 5. We further consider perturbations
due to a weak magnetic field and single-ion spin-orbit interaction.
In the weak-pairing regime of our model, the ground state remains
gapless up to a critical field strength.  The Fermi points of this
critical phase are characterized by a nonzero topological invariant,
hence are stable against weak perturbations. As the field strength
is increased, Fermi points with opposite winding numbers eventually
annihilate with each other and the spectrum acquires a gap above a
critical field. Since the TRS is explicitly broken in the presence
of a uniform magnetic field, the gapped phase represents a
multilayer generalization of the 2D Kitaev model, and is
topologically trivial in three dimensions.

On the other hand, when the sign of magnetic field alternates
between successive honeycomb planes, an effective TRS is restored in
the auxiliary fermionic model, and the corresponding ground state in
the weak-pairing phase is equivalent to a topological superconductor
belonging to symmetry class DIII in Altland-Zirnbauer's
classification. \cite{az} We also show that the quantum ground state
is characterized by an integer winding number, consistent with the
general classification of 3D topological
insulators.\cite{schnyder08}  A physical consequence of the
nontrivial ground-state topology is the appearance of
surface Majorana fermions which remain gapless against perturbations
respecting the discrete symmetries of the Hamiltonian. The specific
multilayer geometry of our model also allows us to analytically
demonstrate the existence of surface Majorana fermions obeying a
Dirac-like Hamiltonian by relating the surface modes to the chiral
edge modes of 2D Kitaev model.

\section{The spin-orbital model}

\begin{figure} [t]
\includegraphics[width=0.8\columnwidth]{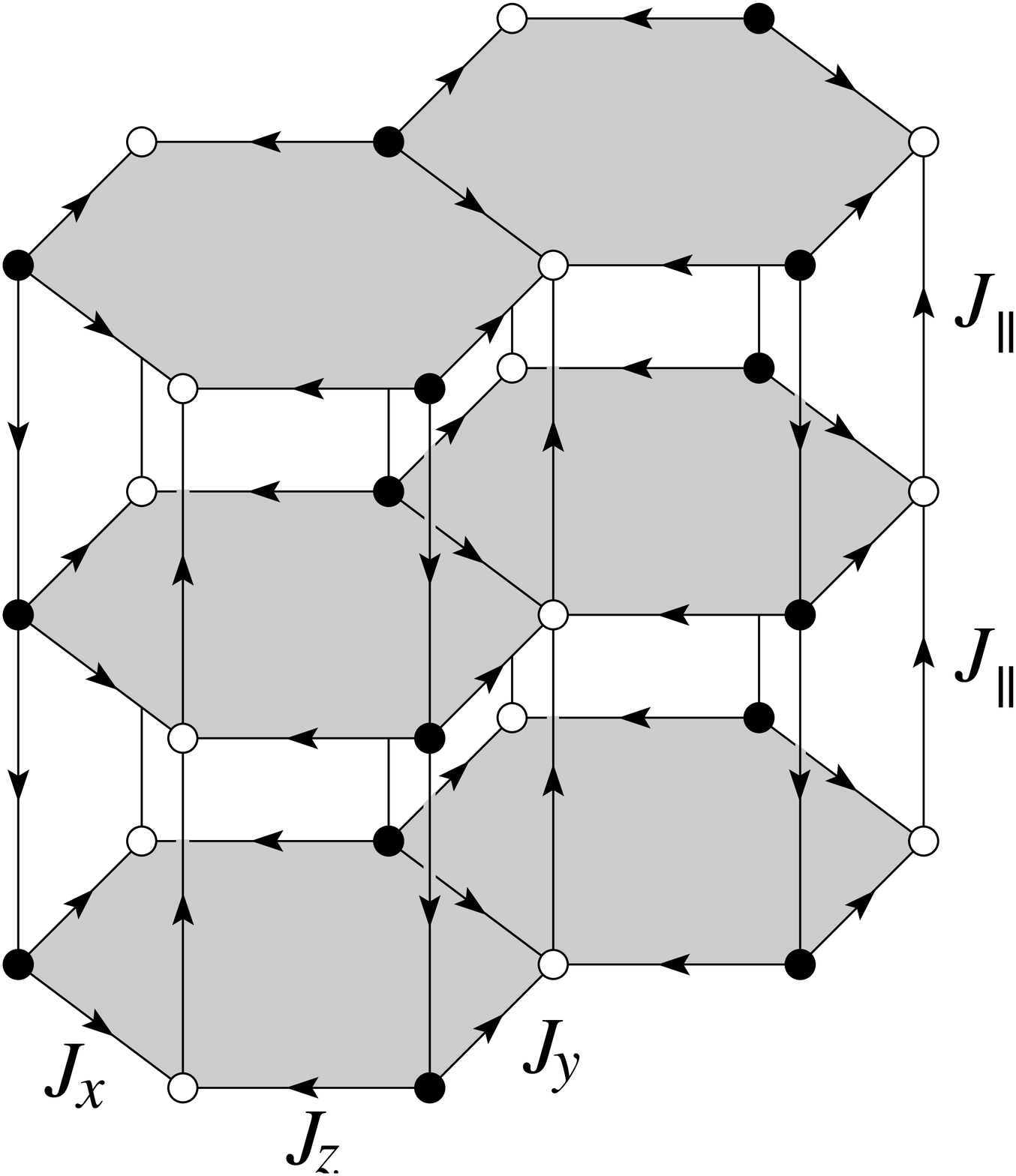}
\caption{\label{fig:honeycomb} A layered honeycomb lattice. The five
distinct nearest-neighbor links are indicated by the corresponding
exchange constant. The open and filled circles denote sites
belonging to the two inequivalent sites of a unit cell. The
primitive vectors of the underlying Bravais lattice are $\mathbf a_1
= (1,0,0)$, $\mathbf a_2 = (\frac{1}{2},\frac{\sqrt{3}}{2},0)$, and
$\mathbf a_3 = (0,0,1)$. An arrow from site $j$ to site $k$ means
the corresponding link variable $u_{jk} = +1$.}
\end{figure}

We study a Kugel-Khomskii-type\cite{kk} spin-orbital model defined
on a layered honeycomb lattice shown in Fig.~\ref{fig:honeycomb}. In
addition to a spin-1/2 degree of freedom, each lattice site has an
extra doublet orbital degeneracy. The spin and orbital (pseudospin)
operators are denoted by Pauli matrices $\sigma^\alpha$ and
$\tau^\alpha$ ($\alpha = x, y, z$), respectively. As in the original
honeycomb Kitaev model, nearest-neighbor links lying on a honeycomb
plane are divided into three types: $x$, $y$, and $z$, depending on
their orientations. The model Hamiltonian is defined as follows:
\begin{eqnarray}
    \label{eq:h}
    \mathcal{H}_0 &=& -\sum_{j}\!^{\,'} \, J_{\parallel}
    \bigl(\tau^x_{j}\tau^x_{j +  z} +
    \tau^y_{j}\tau^y_{j -  z}\bigr) \nonumber \\
    & & -\frac{1}{2} \sum_{j}\sum_{\alpha=x,y,z} J_\alpha\,
    \tau^z_{j}\tau^z_{j+ \bm\delta_\alpha}
    \,\sigma^\alpha_{j}\sigma^\alpha_{j + \bm\delta_\alpha}.
\end{eqnarray}
Here $j$ runs over the lattice sites, $j\pm  z$ denote nearest
neighbors along the two vertical links, and $j + \bm\delta_\alpha$
denotes in-plane nearest neighbor along the link of type $\alpha$.
The prime in the first term indicates that the summation runs over
sites on every second honeycomb layer. The exchange constant is
$J_{\parallel}$ on vertical links, and is $J_\alpha$ on
$\alpha$-links lying on a honeycomb plane (see
Fig.~\ref{fig:honeycomb}). The spin-orbital interaction within each
honeycomb layer resembles the original 2D Kitaev model with spin-1/2
operator $\sigma^\alpha$ replaced by the spin-orbital operator
$\tau^z\sigma^\alpha$. The inter-layer interaction in our model, on
the other hand, involves only orbital operators; the orbital
interaction alternates between the $\tau^x$ and $\tau^y$ types along
successive vertical links. In addition to discrete lattice
symmetries, the Hamiltonian is invariant under a $\pi$ rotation
about $\tau^z$ axis followed by lattice translations along the $z$
direction by one layer.

Since the five spin-orbital operators $\tau^x$, $\tau^y$, and
$\tau^z\sigma^\alpha$ anticommute with each other, they generate a
$4\times 4$ matrix representation of the Clifford algebra. With an
enlarged Hilbert space, one may introduce 6 Majorana fermions $c$
and $b^\mu$ ($\mu = 1,\cdots,5$) such that:
\begin{eqnarray}
    & &\tau^z\sigma^x = i b^1 c, \quad
    \tau^z\sigma^y = i b^2 c, \quad
    \tau^z\sigma^z = i b^3 c, \nonumber \\
    & & \quad\quad\quad\quad
    \tau^x = i b^4 c, \quad
    \tau^y = i b^5 c.
\end{eqnarray}
By denoting these operators as $\Gamma^\mu = i b^\mu c$,
Hamiltonian~(\ref{eq:h}) can be recast into a Kitaev-type
interaction
\begin{eqnarray}
    \mathcal{H}_0 = -\sum_{\mu=1}^5 J_{\mu} \sum_{\mu-\mbox{\scriptsize
    links}}  \Gamma^\mu_j \Gamma^\mu_k,
\end{eqnarray}
with $J_1 = J_x$, $J_2 = J_y$, and $J_3 = J_z$ for links lying on a
honeycomb plane, and $J_4 = J_5 = J_{\parallel}$ along the vertical
links. The six Majorana fermions form an 8-dimensional Hilbert space
which is twice as large as the local physical Hilbert space. This
redundancy can be remedied by demanding the allowed physical states
be eigenstate of gauge operator $D \equiv i c \,\prod_{\mu=1}^5
b^\mu$ with eigenvalue +1. This is also consistent with the identity
$\tau^x\tau^y\tau^z\sigma^x\sigma^y\sigma^z = -1$. We note that the
same $\Gamma$-matrix extension of the Kitaev model on a 2D decorated
square lattice (Shastry-Sutherland lattice) has been studied in Ref.
\onlinecite{wu09}.

Following Kitaev, \cite{kitaev06} we introduce the link operator
$u_{jk} \equiv i b^\mu_j\,b^\mu_k$, where $\mu=\mu_{jk}$ implicitly
depends on the type of link connecting sites $j$ and $k$. The
Hamiltonian then becomes
\begin{eqnarray}
    \label{eq:H-cc}
    \mathcal{H}_0 = \frac{i}{2} \sum_{jk} J_{\mu}u_{jk}\, c_j c_k.
\end{eqnarray}
A remarkable feature of the fermionic Hamiltonian first noted by
Kiatev is that the link operators $u_{jk}$ commute with each other
and with the Hamiltonian. We may thus replace them by their
eigenvalues $u_{jk} = \pm 1$, which act as an emergent $Z_2$ gauge
field. Consequently, for a given choice of $\{u_{jk}\}$, Hamiltonian
(\ref{eq:H-cc}) reduces to a problem of free Majorana fermions with
nearest-neighbor hopping $t_{jk} = J_{\mu} u_{jk}$. However, since
$u_{jk}$ does not commute with operator $D_j$, the spectrum of the
free fermion Hamiltonian depends only on gauge invariant quantities
which are given by the product of link operators around the boundary
of elementary plaquette: $w_p = \prod_{(jk)\in
\partial p} u_{jk}$. As $w_p^2 = 1$, the flux associated with
a given plaquette is also given by a $Z_2$ variable $w_p = \pm 1$.

\begin{figure}
\includegraphics[width=0.95\columnwidth]{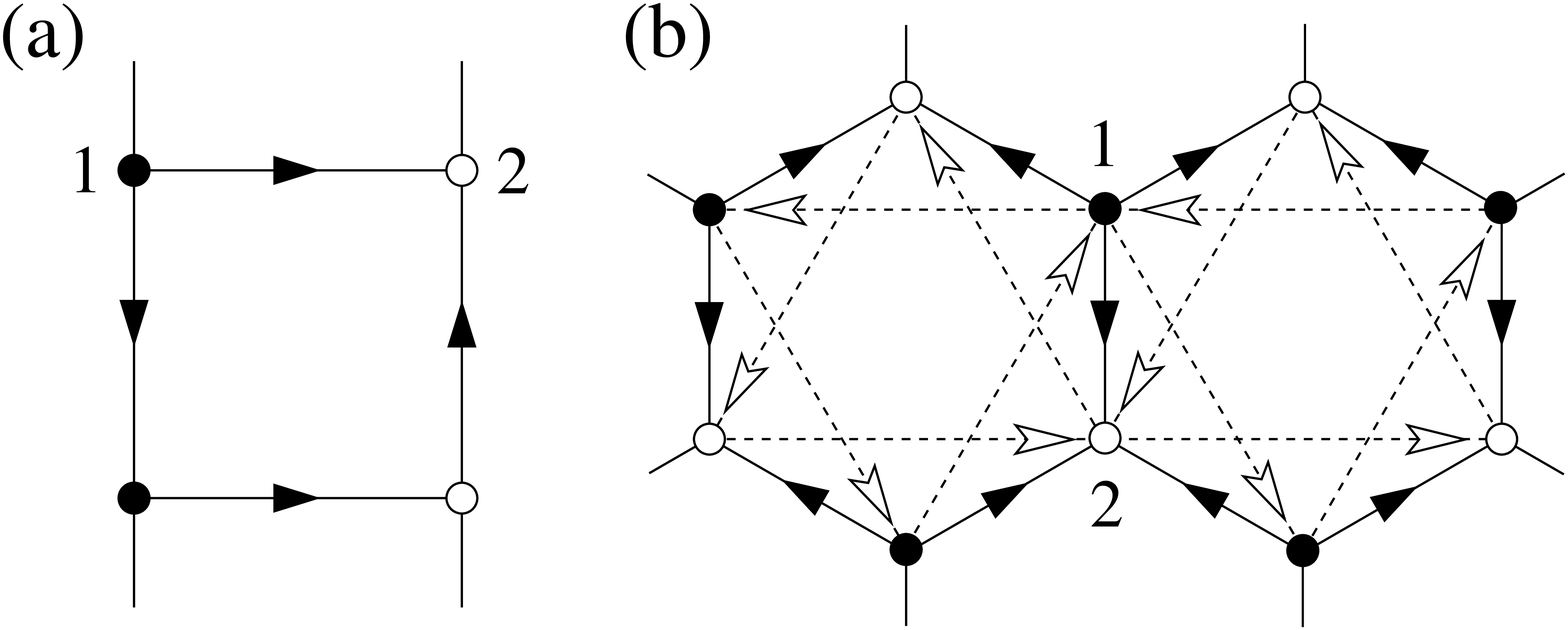}
\caption{\label{fig:flux} Configuration of $Z_2$ gauge fields
$\{u_{jk}\}$ and $\{u'_{jk}\}$ on (a) square and (b) hexagonal
plaquettes; $u_{jk} = 1$ if there is an arrow pointing from site $j$
to site $k$. The numbers 1 and 2 label the two inequivalent sites
within a unit cell.}
\end{figure}

The question remains of what choices of $u_{jk}$, or equivalently
the $Z_2$ fluxes $w_p$, give the lowest ground-state energy of the
free fermion problem. Numerically, we find that such a state is
attained when all hexagons are vortex free while all squares contain
a $\pi$ flux:
\begin{eqnarray}
    w_{\varhexagon} = 1, \quad\quad w_{\Square} = -1,
\end{eqnarray}
consistent with Lieb's theorem. \cite{lieb} A gauge choice which
gives the desired plaquette fluxes without enlarging the unit cell
is shown in Fig.~\ref{fig:flux}. With this specific choice of $Z_2$
gauge field, the fermion spectrum can be obtained analytically using
Fourier transformation.

In the weak-pairing regime to be discussed below, model (\ref{eq:h})
exhibits a gapless phase similar to the B phase of 2D Kitaev model.
In order to explore possible topological insulators emerging from
this critical phase, we consider the following single-ion
perturbations:
\begin{eqnarray}
    \label{eq:h1}
    \mathcal{H}_1 = -\sum_{j} \mathbf h_j\cdot\bm \sigma_j
    + \lambda\sum_{j} \tau^z_{j} \hat\mathbf n\cdot\bm\sigma_j.
\end{eqnarray}
The first term is the Zeeman coupling to an external magnetic field
$\mathbf h_j = \eta_j \mathbf h$, where the phase factor $\eta_j =
\pm 1$ depends only on the $z$ coordinate of spin $j$, i.e. the
field is constant within a given honeycomb plane. As mentioned
previously, we shall consider two special cases in this paper:
uniform field with $\eta_j = 1$ and staggered field with $\eta_j =
(-1)^{z_j}$. Since spins transform as $T \sigma^\alpha T^{-1} =
-\sigma^\alpha$ under time-reversal $T$, the first term above
explicitly breaks TRS.

The second term in Eq.~(\ref{eq:h1}) represents a spin-orbit-like
interaction where $\lambda$ is an effective coupling constant, and
$\hat\mathbf n$ is a unit vector specifying the local anisotropy
axis. Such a coupling arises when the orbital basis $|\tau^z = \pm
1\rangle$ carries a nonzero orbital angular momentum $\mathbf L \sim
\tau^z\hat\mathbf n$ which is parallel or antiparallel to the local
anisotropy axis depending on $\tau^z = +1$ or $-1$, respectively. An
explicit example is given by two degenerate orbitals with $t_{2g}$
symmetry, e.g. $|yz\rangle$ and $|zx\rangle$ in a tetragonal crystal
field. By identifying $|\tau^z = \pm 1 \rangle$ as $|yz\rangle \pm i
|zx \rangle$, respectively, the orbital basis has a nonzero angular
momentum pointing along the symmetry axis of the tetragonal field.

The effects of $\mathcal{H}_1$ can be studied following the
perturbation treatments discussed in Ref.~\onlinecite{kitaev06}.
Essentially, one constructs an effective Hamiltonian acting on the
subspace which is free of vortex-type excitations. The first-order
correction vanishes identically as both single-ion perturbations
create $\pi$ fluxes on hexagons, \cite{baskaran07} whereas the
second-order terms simply modify the nearest-neighbor exchange
constants $J_{\alpha}$. Nontrivial corrections to the fermion
spectrum arise at the third-order perturbation which involves
multiple spin-orbital interactions, e.g.
\[
    (\tau^z_j \sigma^x_j)(\tau^z_k \sigma^y_k)
    \,\sigma^z_l = i (\tau^z_j \sigma^x_j)(\tau^z_k \sigma^y_k)
    (\tau^z_l \sigma^x_l) (\tau^z_l \sigma^y_l).
\]
Such terms introduce an effective second nearest-neighbor hopping
for Majorana fermions
\begin{eqnarray}
    \label{eq:H-cc2}
    \mathcal{H}_1 = \frac{i \kappa}{2} \sum_{jk} \eta \, u'_{jk} c_j c_k,
\end{eqnarray}
where $\kappa \sim \lambda^2 h/J^2$ and $\eta = \pm 1$ is a constant
within the plane containing sites $j$ and $k$. The additional
second-neighbor $Z_2$ field $u'_{jk} = -u'_{kj}$ is shown by the
dashed line in Fig.~\ref{fig:flux}. Depending on the direction of
$\hat\mathbf n$ and $\mathbf h$, the hopping amplitude $\kappa$
could take different values along inequivalent second-neighbor
links. As the main purpose of this term is to introduce a spectral
gap, to avoid unnecessary complications, we shall assume the
symmetric case in the following discussion.

\section{Uniform magnetic field}

\label{sec:uniform-h}

We first discuss the ground state in the presence of a uniform
magnetic field, i.e. $\eta_j = +1$ for all layers. As will be
discussed below, the gapped phase at large fields is characterized
by a $\pi_2$ Chern number corresponding to a multi-layer
generalization of the 2D Kitaev model, and is topologically trivial
in 3D (which generally is characterized by $\pi_3$ homotopy groups).
This is mainly because the TRS essential to the existence of 3D
topological insulators is explicitly broken by the uniform field.
However, the critical phase in the case of small fields is
interesting in itself and is similar to the gapless A-phase of
$^3$He discussed in Ref.~\onlinecite{volovik}; the gaplessness of
both phases are topologically protected.

Since the original unit cell of the lattice is preserved by the
$Z_2$ fields $u_{jk}$ and $u'_{jk}$ shown in Fig.~\ref{fig:flux}, we
express the site index as $j = (\mathbf r, s)$, where $\mathbf r$
denotes the position of the unit cell, and $s=1,2$ indicates the two
inequivalent sites in a unit cell. With Fourier transformation
$a_{\mathbf k,s} = \sum_{\bf r} c_{\mathbf r, s}\, e^{-i\mathbf
k\cdot(\mathbf r + \mathbf d_s)}/\sqrt{2N}$, where $\mathbf d_s$ is
a basis vector and $N$ is the number of unit cells, the Hamiltonian
becomes $\mathcal{H} = \mathcal{H}_0 + \mathcal{H}_1 =\frac{1}{2}
\sum_{\mathbf k}\,\Psi^\dagger_{\mathbf k}\,H(\mathbf
k)\,\Psi_{\mathbf k} $, with $\Psi_{\mathbf k} = (a_{\mathbf k,1},
\, a_{\mathbf k, 2})^T$ and
\begin{eqnarray}
    \label{eq:Hu}
    H(\mathbf k) &=& \left(\begin{array}{cc}
    g(k_z)+\Delta(\mathbf k_\perp) & -i f(\mathbf k_\perp) \\
    i f(\mathbf k_\perp)^* & -g(k_z)-\Delta(\mathbf k_\perp)
    \end{array}\right), \\
    & = & {\rm Im} f(\mathbf k_\perp)\,\tau^x + {\rm Re} f(\mathbf k_\perp)\,\tau^y +
    \bigl[g(k_z)\!+\!\Delta(\mathbf k_\perp)\bigr]\,\tau^z. \nonumber
\end{eqnarray}
For convenience, we have defined the following functions:
\begin{eqnarray}
    g(k_z) = 4 J_\parallel  \sin k_z, \quad
    f(\mathbf k_\perp) = 2 \sum_{\alpha=1}^3 J_\alpha\,e^{i\mathbf
    k_{\perp}\cdot \bm \delta_\alpha}, \\ \nonumber
    \Delta(\mathbf k_\perp) =
    8\kappa\,\sin\frac{k_x}{2}\,\Bigl(\cos\frac{\sqrt{3}k_y}{2}-\cos\frac{k_x}{2}\Bigr).
    \quad
\end{eqnarray}
The three vectors $\bm \delta_{1,2} = (\frac{\pm
1}{2},\frac{-1}{2\sqrt{3}})$, and $\bm \delta_{3} =
(0,\frac{1}{\sqrt{3}})$ connect nearest-neighbor sites in a
honeycomb layer. In the following we shall focus on the emergent
free fermion problem. The Pauli matrices $\tau^\mu$ appearing in the
single-particle Hamiltonian now act on the sublattice index $s$ (not
to be confused with orbital pseudospins).

The hermitian matrix $H(\mathbf k)$ in Eq.~(\ref{eq:Hu}) also
satisfies
\begin{eqnarray}
    \label{eq:symD}
    H^T(-\mathbf k) = - H(\mathbf k),
\end{eqnarray}
which is the defining property of symmetry class~D in
Altland-Zirnbauer's classification. \cite{az} As 3D insulators in
this class is topologically trivial, \cite{schnyder08} the gapped
phase of Eq.~(\ref{eq:Hu}) represents a trivial multilayer
generalization of the 2D Kitaev model. Nonetheless, for small
magnetic fields such that the second-neighbor hopping $\kappa \sim
\lambda^2 h$ is below a critical value $\kappa_c$, the fermion
spectrum remains gapless in the weak-paring regime of the model. The
corresponding critical phase is characterized by topologically
protected Fermi points as we shall discuss below.

Diagonalizing Hamiltonian (\ref{eq:Hu}) yields a spectrum:
\begin{eqnarray}
    \label{eq:e1}
    \epsilon_{\pm}(\mathbf k) = \pm \sqrt{\bigl[g(k_z) + \Delta(\mathbf
    k_\perp)\bigr]^2 + |f(\mathbf k_\perp)|^2}.
\end{eqnarray}
When one of the in-plane coupling is much larger than the other two,
e.g. $J_z \gg J_x, J_y$, there is no solution to equation $f(\mathbf
k_\perp) = 0$, and the spectrum is always gapped irrespective of the
applied magnetic field. This phase corresponds to the A phase of the
2D Kitaev model, \cite{kitaev06} and is referred to in the following
as the strong-pairing phase based on analogy with the $p$-wave
topological superconductors. \cite{read00} On the other hand, in the
weak-pairing regime of the model where the three in-plane couplings
satisfy the triangle inequalities, \cite{kitaev06} the model
displays a possible gapless phase as two solutions $\mathbf k^*_\perp$
exist for the equation $f(\mathbf k_\perp) = 0$.
To be specific, we now concentrate on the symmetric case $J_x = J_y =
J_z \equiv J$, where zeros of $f(\mathbf k_\perp)$ are at the
corners of the 2D hexagonal Brillouin zone $\mathbf k^*_{\perp} =
(\pm \frac{4\pi}{3}, 0)$.

In contrast to the 2D Kitaev model where applying a magnetic field
immediately opens an energy gap, \cite{kitaev06} spectrum
(\ref{eq:e1}) of the 3D model remains gapless when $\kappa$ is less
than a critical strength $|\kappa| \leq \kappa_c.$ For symmetric
in-plane couplings, we find $\kappa_c \equiv 2J_\parallel/3\sqrt{3}$
and the nodes of the spectrum $\epsilon_{\pm}(\mathbf k)$ are
located at
\begin{eqnarray}
    \label{eq:fermipts}
    \begin{array}{llcl}
    \quad &  \mathbf k^*_{1,+} = \bigl(\frac{4\pi}{3}, 0, -\pi+\xi\bigr),
    &  &
    \mathbf k^*_{1,-} = \bigl(\frac{4\pi}{3}, 0, -\xi\bigr),
    \\
    \quad & \mathbf k^*_{2,+} = \bigl(-\frac{4\pi}{3}, 0, \xi\bigr),
    &  &
    \mathbf k^*_{2,-} = \bigl(-\frac{4\pi}{3}, 0, \pi-\xi\bigr),
    \end{array} \quad
\end{eqnarray}
where $\xi = \arcsin(\kappa/\kappa_c)$. These Fermi points are
topologically protected and are robust against weak perturbations.
To see this, we first rewrite Hamiltonian (\ref{eq:Hu}) as
$H(\mathbf k) = \varepsilon_+(\mathbf k)\,\hat{\mathbf m}(\mathbf
k)\!\cdot\!\bm\tau$, where $\hat\mathbf m(\mathbf k)$ is a unit
vector. A topological invariant characterizing the singularities of
the spectrum is given by the winding number of mappings from a
sphere $S^2$ enclosing the Fermi point $\mathbf k^*$ to the 2-sphere
of the unit vector $\hat{\mathbf m}$: \cite{volovik}
\begin{eqnarray}
    \label{eq:nu1}
    \nu = \int_{S^2} \frac{d^2 A_\rho}{8\pi} \epsilon^{\mu\nu\rho}\,
    \hat{\mathbf m}\cdot\bigl(\partial_\mu\hat{\mathbf m}\times\partial_\nu\hat{\mathbf m}\bigr).
\end{eqnarray}
The above definition corresponds to the second homotopy group
$\pi_2(S^2) = Z$, which characterizes point defects in an $O(3)$
spin field.\cite{mermin79}

Examples of topologically nontrivial Fermi points are given by the
spectra of Weyl Hamiltonian describing a massless spin-1/2 particle:
$H_{\rm Weyl} = \pm ic\tau^\mu\partial_\mu$, where $c$ is the speed
of light. \cite{volovik} The plus and minus signs refer to left and
right-handed particles, respectively. Using Eq.~(\ref{eq:nu1}) the
winding number of Weyl spinors can be readily computed, resulting
$\nu = \pm 1$ for right and left-handed particles, respectively.
Take left-handed Weyl spinor for example, the expectation value of
its spin is parallel to its momentum: $\langle \bm \tau \rangle
\parallel \mathbf p$. In the ground state with filled
negative-energy states, Fermi point with $\nu = 1$ thus looks like a
magnetic monopole (a hedgehog) in momentum space. These singular
points are robust in the sense that it is impossible to continuously
deform a hedgehog into a uniform spin configuration (corresponding
to the trivial case of $\nu = 0$).

\begin{figure}
\includegraphics[width=0.8\columnwidth]{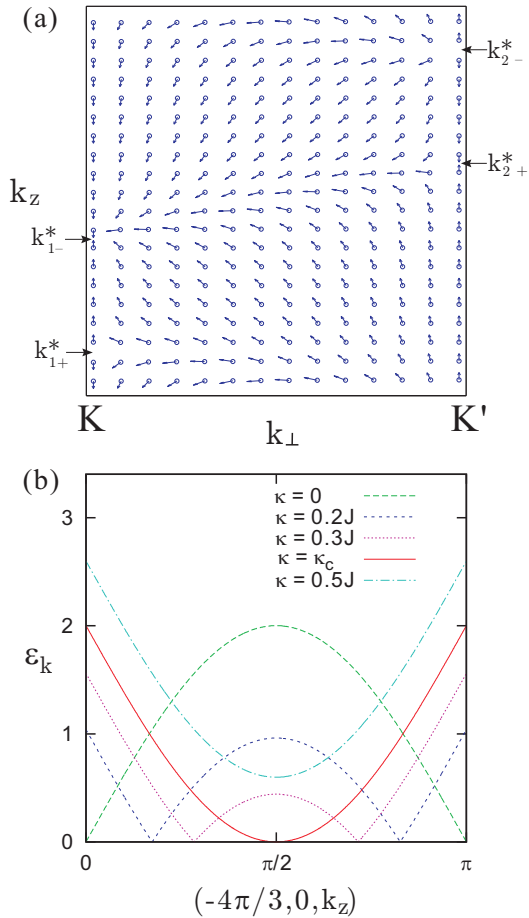}
\caption{\label{fig:k-top} (Color online) (a) Projection of
ground-state pseudospin $\langle \bm\tau\rangle = -\hat{\mathbf
m}(\mathbf k)$ on a face of the Brillouin zone. The two vertical
edges of the face correspond to $K = (\frac{4\pi}{3},0, k_z)$ and
$K'=(\frac{2\pi}{3},\frac{2\pi}{\sqrt{3}},k_z) \equiv
(-\frac{4\pi}{3},0,k_z)$. The location of the four topologically
non-trivial Fermi points are indicated by black arrows. (b)
Dispersion along the $K'$ edge of Brillouin zone for various $\kappa
\sim \lambda^2 h$. Upon increasing magnetic field $h$, the two Fermi
points carrying opposite winding numbers move in opposite directions
along the $K'$ edge. At a critical field ($\kappa = \kappa_c$), the
two singularities merge  to form a new Fermi point with trivial
winding number $\nu = 0$ at $(-\frac{4\pi}{3}, 0, \frac{\pi}{2})$.
The spectrum becomes gapped above $\kappa_c$.}
\end{figure}

To compute the winding number of Fermi points in our 3D model,
we expand Hamiltonian (\ref{eq:Hu}) around the
singular points, e.g.
\begin{eqnarray}
    \label{eq:hw}
    H(\mathbf k^*_{1,\pm}+\mathbf p) = - v_\perp p_y\tau^x - v_\perp
    p_x\tau^y \mp v_\parallel p_z \tau^z,
\end{eqnarray}
where the Fermi velocities are $v_\perp = \sqrt{3}J$ and
$v_\parallel\!=\! 2 (4 J_\parallel^2 - 27 \kappa^2 )^{1/2}$. The
dispersion around these points has a conic
singularity:~$\varepsilon_\pm\!\approx\! \pm \sqrt{(v_\perp \mathbf
p_\perp)^2 + (v_\parallel p_z)^2}$. After rotation and
mirror-inversion, Eq.~(\ref{eq:hw}) is essentially equivalent to the
momentum-space Weyl Hamiltonian discussed above. The four Fermi
points (\ref{eq:fermipts}) form two pairs (labeled by $r = 1,2$) of
singularities with opposite winding numbers $\nu = \pm 1$. The
ground-state configuration of pseudospin $\langle \bm\tau\rangle =
-\hat{\mathbf m}(\mathbf k)$ projected onto a face of the Brillouin
zone is shown in Fig.~\ref{fig:k-top}(a). The two pairs of
singularities are located at the two inequivalent edges $K$ and $K'$
of the 3D Brillouin zone. Instead of a monopole-like configuration,
vector field $\hat\mathbf m(\mathbf k)$ in the vicinity of these
Fermi points has a saddle-point-like singularity.

Although these Fermi points are topologically protected due to their
nonzero winding numbers, a spectral gap can still be induced through
mutual annihilation of Fermi points carrying opposite topological
charges. This process is illustrated in Fig.~\ref{fig:k-top}. Take
for example the two Fermi points on the $K'$-edge of the Brillouin
zone. Upon increasing magnetic field $h$, the two singularities
characterized by winding numbers $\nu = \pm 1$ move toward each
other along the $K'$ edge, and eventually merge to form a new Fermi
point when $\kappa$ reaches the critical $\kappa_c$. Because of the
conservation of topological charge, the new Fermi point has a
winding number $\nu = 0$, hence is topologically trivial. Above the
critical field, the spectrum is fully gapped.

It is also interesting to understand the critical B phase of
Kitaev's original 2D model from the perspective of topological
winding number. As discussed in Ref.~\onlinecite{kitaev06}, the
gaplessness of B phase is protected by TRS on the bipartite
honeycomb lattice. This discrete symmetry essentially forces the
unit vector $\hat\mathbf m(\mathbf k_\perp)$ to lie in the $xy$
plane as $\tau^z$ terms in Eq.~(\ref{eq:Hu}) is not allowed by TRS.
For a planar unit vector $\hat\mathbf m$ whose tip lies on a circle
$S^1$, the singularities are characterized by an integer topological
invariant, also known as the vortex winding number, corresponding to
$\pi_1(S^1) = Z$. \cite{mermin79} The two Fermi poins in the B phase
of Kitaev's model can be viewed as vortices carrying opposite
winding numbers $\nu = \pm 1$, respectively. In the presence of
perturbations breaking the TRS, the vector $\hat\mathbf m$ now lives
on a 2-sphere. Since $\pi_1(S^2) = 0$, singularities of the spectrum
are then topologically trivial ($\pi_2$ characterization of 2D
singularities is not well defined).

The gapped phase above the critical $\kappa_c$ represents a trivial
generalization of the 2D Kitaev model in much the same way as the
multilayer generalization of the IQH state.  The topological
properties of such systems are characterized by three spectral Chern
numbers. \cite{kohmoto92} In our case, the nonzero topological
invariant is given by the winding number of vector field
$\hat{\mathbf m}(\mathbf k_\perp; k_z)$ which maps the in-plane
hexagonal Brillouin zone (a 2-torus) to a unit 2-sphere:
\begin{eqnarray}
    \label{eq:nu1b}
    \nu = \frac{1}{4\pi}\int dk_x dk_y\,\hat{\mathbf
    m}\cdot\bigl(\partial_x \hat{\mathbf
    m}\times\partial_y\hat{\mathbf m}\bigr)
    = \frac{\kappa}{|\kappa|}.
\end{eqnarray}
By treating  $k_z$ as a parameter, Hamiltonian (\ref{eq:Hu}) has
exactly the same form as that of the 2D Kitaev model. The first
Chern number (\ref{eq:nu1b}) of the corresponding `2D' Hamiltonian
$H_{k_z}(\mathbf k_\perp)$ is given by $\nu = {\rm
sgn}\,\Delta_{k_z}$, \cite{kitaev06} where $\Delta_{k_z} =
6\sqrt{3}(\kappa + \kappa_c\,\sin k_z)$ is the effective gap
parameter at corner $K$ of the hexagonal Brillouin zone. Therefore,
as long as $|\kappa| > \kappa_c$, hence the system remains gapped,
the winding number  $\nu  = {\rm sgn}\kappa = \pm 1$ is independent
of $k_z$, and the whole Brillouin zone is characterized by the same
chirality.

\section{Staggered magnetic field}

\label{sec:stagger-h}

The gapped phase discussed in the previous section is topologically
trivial due to the absence of TRS. As discussed in
Ref.~\onlinecite{schnyder08}, TRS is a prerequisite for the
existence of 3D topological insulators. In this section, we show
that one can introduce a gap to the fermion spectrum, while at the
same time preserving an effective TRS, by subjecting the system to a
staggered magnetic field, i.e. $\eta_j = (-1)^{z_j}$. Because the
sign of the field alternates between successive honeycomb layers, a
discrete symmetry emerges in our model system as
Hamiltonian~(\ref{eq:h1}) is invariant under time-reversal $T$
followed by a lattice translation along $z$ axis. This additional
symmetry manifests itself as a TRS in the auxiliary Majorana hopping
problem. The mechanism proposed here is similar to Haldane's model
of realizing 2D quantum Hall effect without a net magnetic flux
through the unit cells. \cite{haldane88}

Due to the staggered field, the sign of second nearest-neighbor
hopping $u'_{jk}$ also alternates between successive honeycomb
planes. This results in a staggered gap function $ \pm
\Delta(\mathbf k_{\perp})$, and a doubled unit cell along $z$
direction. We denote the fermion annihilation operators on the even
and odd layers as $a_{\mathbf r, s}$ and $b_{\mathbf r, s}$,
respectively, where subscript $s = 1, 2$ refers to the two inequivalent
sites within a honeycomb plane. The Fourier-transformed Hamiltonian
then reads $\mathcal{H} = \frac{1}{2}\sum_{\mathbf k}
\Psi^\dagger_{\mathbf k} H(\mathbf k) \Psi_{\mathbf k}$, with
\begin{eqnarray}
    \label{eq:Hs}
    & &H(\mathbf k) = \left(\begin{array}{cccc}
    \Delta(\mathbf k_{\perp}) & g(k_z) & -i f(\mathbf k_{\perp}) & 0
    \\
    g(k_z) & -\Delta(\mathbf k_{\perp}) & 0 &
    -i f(\mathbf k_{\perp}) \\
    i f(\mathbf k_{\perp})^* & 0 & -\Delta(\mathbf k_{\perp}) & -g(k_z) \\
    0 & i f(\mathbf k_{\perp})^* & -g(k_z) & \Delta(\mathbf k_\perp)
    \end{array}\right) \quad\,\,\,\, \\ \nonumber
    \\
    & & \,\, =  \Delta(\mathbf k_{\perp}) \tau^z \sigma^z\!+\! g(k_z)
    \tau^z\sigma^x\! +\! {\rm Im}f(\mathbf k_{\perp})\tau^x \! +\! {\rm Re}
    f(\mathbf k_{\perp})\tau^y, \nonumber
\end{eqnarray}
and $\Psi_{\mathbf k} = \bigl(a_{\mathbf k,1}, b_{\mathbf k_,1},
a_{\mathbf k,2},  b_{\mathbf k, 2}\bigr)^T$. The two sets of Pauli
matrices $\tau^\mu$ and $\sigma^\mu$ now act on the sublattice
$(1,2)$ and even-odd $(a,b)$ indices, respectively.

In addition to symmetry relation (\ref{eq:symD}) shared by
Hamiltonians describing free Majorana fermions, the hermitian matrix
(\ref{eq:Hs}) also satisfies
\begin{eqnarray}
    \label{eq:trs}
    \tau^z (i\sigma^y)\, H^T(\mathbf k)\, (-i\sigma^y)\, \tau^z =
    H(-\mathbf k),
\end{eqnarray}
stemming from the generalized TRS. The extra $\tau^z$ factor can be
gauged away by a $\pi/2$ rotation about the $\tau^z$ axis.
Eq.~(\ref{eq:trs}) defines the symmetry property of DIII
Hamiltonians in Altland-Zirnbauer's classification. \cite{az} As
discussed in Ref.~\onlinecite{schnyder08}, a topological invariant
can be defined for Hamiltonians in this symmetry class based on the
block off-diagonal representation of the Hamiltonian, or more
precisely, of the spectral projection operator. In fact, the same
definition can be applied to classes of Hamiltonians which possess some
form of chiral symmetry arising from either the sublattice symmetry or a
combination of particle-hole and time-reversal symmetries.
\cite{schnyder08}

To compute the topological winding number of our model, we first
bring Hamiltonian (\ref{eq:Hs}) into a block off-diagonal form
through a series of unitary transformations. First, noting that the
layered honeycomb lattice is bipartite in which nearest neighbors
of one sublattice belong to the other one, we regroup fermions of
the same sublattice into a block, e.g. $a_{\mathbf k,1}$ and
$b_{\mathbf k, 2}$. Mathematically this is achieved by interchanging
the 2nd and 4th entries of $\Psi_{\mathbf k}$, the transformed
Hamiltonian becomes
\begin{eqnarray}
    \label{eq:Hd1}
    H(\mathbf k) \to \alpha^x {\rm Im} f(\mathbf k_{\perp})
    + \alpha^y {\rm Re} f(\mathbf k_{\perp})
    + \alpha^z g(k_z)
    + \beta \Delta(\mathbf k_{\perp}),\quad\,
\end{eqnarray}
where $\alpha^\mu$ and $\beta$ given by
\begin{eqnarray}
    \alpha^\mu = \gamma^\mu = \tau^x \sigma^\mu,  \quad
    \beta = \gamma^0 = \tau^z,
\end{eqnarray}
are the standard Dirac matrices. The second part of the unitary
transformation is a $\pi/2$ rotation about the new $\tau^x$ axis,
which transforms $\tau^z \to \tau^y$, hence
\begin{eqnarray}
    \label{eq:HDIII}
    H(\mathbf k) \to \left(\begin{array}{cc}
    0 & D(\mathbf k) \\
    D^\dagger(\mathbf k) & 0 \end{array}\right),
\end{eqnarray}
where the upper right block is
\begin{eqnarray}
    \label{eq:D1}
    D(\mathbf k) = \left(\begin{array}{cc}
    g(k_z) - i\Delta(\mathbf k_{\perp}) & -if(\mathbf k_{\perp}) \\
    if(\mathbf k_{\perp})^* & -g(k_z) - i \Delta(\mathbf k_{\perp})
    \end{array}\right).
\end{eqnarray}
Noting that $f(-\mathbf k_{\perp}) = f(\mathbf k_{\perp})^*$, and
$\Delta(\mathbf k_{\perp})$, $g(k_z)$ are odd functions of $\mathbf
k$, matrix (\ref{eq:D1}) satisfies a symmetry $D^T(\mathbf k) =
-D(-\mathbf k)$. The block off-diagonal form of the Hamiltonian
implies that $\epsilon(\mathbf k)^2 = D^\dagger(\mathbf k) D(\mathbf
k)$, which gives rise to a fermion spectrum
\begin{eqnarray}
    \label{eq:ek2}
    \epsilon_{\pm}(\mathbf k) = \pm \sqrt{\Delta(\mathbf
    k_{\perp})^2 + g(k_z)^2 + \bigl|f(\mathbf k_{\perp})\bigr|^2}.
\end{eqnarray}
Due to the effective TRS (\ref{eq:trs}), the spectrum is double
degenerate at each wavevector $\mathbf k$, except at possible Fermi
points. It is worth noting that, contrary to the uniform field case,
the staggered field immediately opens an energy gap to the spectrum
in the weak-pairing phase of the model.

The topological properties of the quantum ground state (occupied
Bloch states) in the gapped phase is captured by the following $Q$
matrix: \cite{schnyder08}
\begin{eqnarray}
    Q(\mathbf k) = \left(\begin{array}{cc}
    0 & q(\mathbf k) \\
    q^\dagger(\mathbf k) & 0 \end{array}\right),
    \quad
    q(\mathbf k) = \frac{D(\mathbf k)}{\epsilon_+(\mathbf k)}.
\end{eqnarray}
In fact $Q(\mathbf k)$ represents a `simplified' Hamiltonian
obtained by assigning an energy $\epsilon = - 1$ to all occupied
states and $\epsilon = +1$ to all empty bands of Hamiltonian
$H(\mathbf k)$. \cite{schnyder08,qi09} As long as the system is in the same gapped phase,
one can continuously deform the model such that $H(\mathbf k)$
gradually transforms to the simplified form $Q(\mathbf k)$. Not
surprisingly, the $Q$ matrix is related to the spectral operator via
$Q(\mathbf k) = 1- 2 P(\mathbf k)$. \cite{schnyder08}

It is easy to see that the block matrix $q$ satisfies
\begin{eqnarray}
    q^\dagger(\mathbf k)\, q(\mathbf k) =  1,
    \quad q^T(\mathbf k) = - q(-\mathbf k),
\end{eqnarray}
as expected for a DIII class Hamiltonian. The topological invariant
characterizing the ground state is defined as the integer winding
number of mapping $q: T^3 \to U(2)$,
\cite{schnyder08,volovik}
\begin{eqnarray}
    \label{eq:nu2}
    \nu = \int \frac{d^3 k}{24\pi^2} \epsilon^{\mu\nu\rho} {\rm
    tr}\bigl[(q^{-1} \partial_{\mu} q)(q^{-1} \partial_{\nu} q)
    (q^{-1} \partial_{\rho} q)\bigr],
\end{eqnarray}
where the integral is over the three-dimensional Brillouin zone,
which is essentially a 3-torus $T^3$. For DIII class
Hamiltonians, the winding number $\nu$ can take on an arbitrary
integer, each labels a unique topological class of the quantum
ground state.

\begin{figure} [t]
\includegraphics[width=0.9\columnwidth]{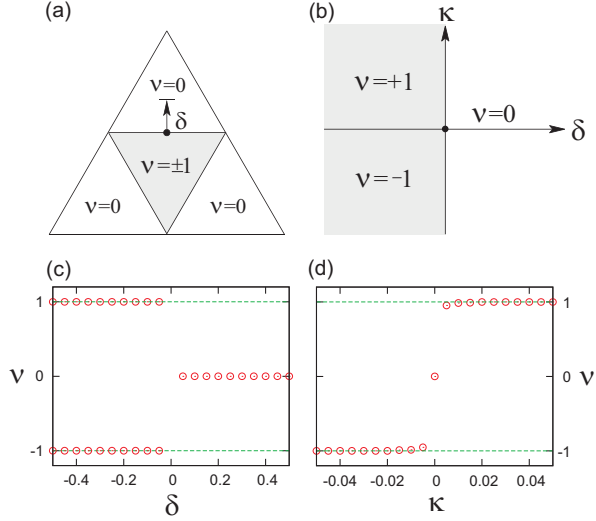}
\caption{\label{fig:phase} (Color online) (a) Phase diagram in the
plane of $J_x + J_y + J_z = $ const. The shaded triangle corresponds
to the region where the in-plane couplings satisfy the triangle
inequalities. In the absence of staggered field $\kappa = 0$, the
shaded area represents a critical phase with zero energy gap. We
define a parameter $\delta = 2 J_x - J_z$ along the $J_x = J_y$ line
specifying the distance from the phase boundary. (b) Phase diagram
as a function of distance $\delta$ and field strength $\kappa$.
Panels (c) and (d) Numerical evaluation of the winding number $\nu$
as a function of $\delta$ and $\kappa$, respectively. The field
strength $\kappa = \pm 0.02 J_z$ in the calculation of panel (c),
whereas we set $J_x = J_y = J_z \equiv J$ and $\kappa$ is measured
in units of $J$ in (d).}
\end{figure}

The value of the topological winding number $\nu$ can only change in
the presence of a quantum phase transition, which is usually
accompanied by a vanishing bulk gap. In our case, noting that $g(0)
= 0$, a prerequisite for spectrum (\ref{eq:ek2}) to be gapless is
that the in-plane couplings satisfy the triangle inequalities $|J_z|
\le |J_x|+|J_y|$, and so on [see Fig.~\ref{fig:phase}(a)], such that
solutions exist for $f(\mathbf k_{\perp}) = 0$ (the weak-paring
regime of the model). In the absence of magnetic field $\kappa = h =
0$, the triangle inequalities thus define a critical phase with
gapless fermionic excitations. For convenience, we use $\delta$ to
denote the `distance' from the boundary of the critical phase. For
example $\delta \equiv 2 J_x - J_z$ along the $J_x = J_y$ line shown
in Fig.~\ref{fig:phase}(a). We numerically compute the winding
number $\nu$ using definition (\ref{eq:nu2}) for various gapped
phases of the model Hamiltonian; the resulting phase diagram as a
function of $\delta$ and $\kappa$ is summarized in
Fig.~\ref{fig:phase}(b).

At the phase boundary defined by $\kappa = 0$ and $\delta = 0$, the
system undergoes a quantum phase transition of a topological nature.
Fig.~\ref{fig:phase}(c) shows numerical evaluation of the winding
number $\nu$ as a function of $\delta = 2 J_x - J_z$ in the presence
of a staggered field such that $|\kappa| = 0.02 J_z$. Depending on
the sign of $\kappa$, the winding number jumps from $\nu = 0$ to
$\nu = \pm 1$ when $\delta$ crosses the phase boundary from the
topologically trivial phase (corresponding to the A phase in 2D
Kitaev model). On the other hand, for systems inside the critical
phase ($\delta < 0$), the winding number jumps from $\nu = -1$ to
$\nu = +1$ as $\kappa$ changes sign.

The nontrivial winding number of the quantum ground state can also
be understood from the topological properties of the singular (Dirac)
points in the fermion spectrum. To simplify the discussion, we focus
on the symmetric case $J_x = J_y = J_z \equiv J$, where zeros of
spectrum (\ref{eq:ek2}) in the $\kappa \to 0$ limit are at
\begin{eqnarray}
    \label{eq:dirac-pts}
    \begin{array}{lcl}
    \mathbf k^*_1 = (\frac{4\pi}{3},0,0), & &
    \mathbf k^*_2 = (-\frac{4\pi}{3}, 0, 0).
    \end{array}
\end{eqnarray}
In the vicinity of these two points, the fermions obey a Dirac-like
Hamiltonian [c.f. Eq. (\ref{eq:Hd1})]
\begin{eqnarray}
    \label{eq:Hc1}
    H(\mathbf k^* + \mathbf p) = \alpha^\mu \tilde p_\mu + m \beta,
\end{eqnarray}
where the scaled momentum and mass term are
\begin{eqnarray}
    \begin{array}{c}
    \tilde \mathbf p_{\perp} = \sqrt{3}J (p_y, \mp
    p_x), \quad \tilde p_z = 4J_{\perp} p_z, \\ \\
    m=\Delta(\mathbf k^*_s) = \pm 6\sqrt{3}\kappa.
    \end{array}
\end{eqnarray}
The plus and minus signs in the expression of $m$ correspond to
$\mathbf k^*_1$ and $\mathbf k^*_2$, respectively. The spectrum of
(\ref{eq:Hc1}) given by $\epsilon_{\pm} = \pm \sqrt{\tilde p^2 +
m^2}$ is two-fold degenerate at each $\mathbf p$. As discussed
above, a $\pi/2$ rotation about $\tau^x$ brings the Dirac mass into
a chiral mass term, i.e. $\beta \to i\gamma^5\beta$. The transformed
Hamiltonian $H = \alpha_\mu \tilde p_\mu - i \beta \gamma^5 m$ is in
a block off-diagonal form with $D = \tilde\mathbf p\cdot\bm\sigma -
i m$. The $q$ matrix of the spectral projector is then given by
\begin{eqnarray}
    \label{eq:qc1}
    q(\mathbf k^* + \mathbf p) = \frac{\tilde\mathbf p\cdot
    \bm\sigma - i m}{\sqrt{\tilde p^2 + m^2}}.
\end{eqnarray}
Eqs. (\ref{eq:Hc1}) and (\ref{eq:qc1}) can be viewed as a continuum
approximation to the low-energy physics of the model system. A direct
evaluation using Eq. (\ref{eq:nu2}) yields a winding number
\begin{eqnarray}
    \nu = \pm\frac{1}{2}\frac{m}{|m|},
\end{eqnarray}
with $\pm$ sign refering to $\mathbf k^*_1$ and $\mathbf k^*_2$
points, respectively. Note that when evaluating $\nu$ using the
continuum description, we have extended the domain of integration to
a 3-sphere in Eq.~(\ref{eq:nu2}). The appearance of a half-integer
$\nu$ is an artifact of the continuum description; the winding
number is modified once contributions from high-energy Bloch states
(away from the Dirac point in the Brillouin zone) are properly
included. Since the mass term has opposite sign at the two Dirac
points, $m(\mathbf k^*_1) = - m(\mathbf k^*_2)$, we obtain $\nu_1 =
\nu_2 = \frac{1}{2} {\rm sgn}\,\kappa$. Interestingly, the sum of
these two winding numbers $\nu = \nu_1 + \nu_2$ reproduces the
topological invariant of the lattice model.

As recently pointed out in Ref.~\onlinecite{beri09}, stable Fermi
lines generally appears in 3D topological superconductors described
by Hamiltonians belonging to classes CI or DIII. An extended phase
diagram of a lattice CI  model \cite{schnyder09b} indeed shows
regions of gapless phase with topologically stable Fermi lines.
\cite{beri09} Here we show that such Fermi lines are also possible
in our DIII Hamiltonian (\ref{eq:Hs}).

To this end, we consider perturbations which break the sublattice
symmetry by introducing different second-neighbor hoppings $\kappa$
on the two inequivalent sites of the lattice: $\kappa_{1,2} = \kappa
\pm \delta\kappa$. In the vicinity of the Dirac points
(\ref{eq:dirac-pts}), the asymmetric coupling results in an
additional term of the off-diagonal block matrix
\begin{eqnarray}
    D =\tilde\mathbf p\cdot\bm\sigma - i m - i \mu \sigma^z,
    \quad\quad \mu = 6\sqrt{3}\,\delta\kappa.
\end{eqnarray}
The corresponding spectrum $\epsilon^2 = \tilde p_z^2 + \bigl(\mu -
\sqrt{m^2 + \tilde p_{\perp}^2}\bigr)^2$ has a Fermi line specified
by
\begin{eqnarray}
    \tilde p_{\perp} = \sqrt{\mu^2 - m^2},
    \quad\quad \tilde p_z = 0,
\end{eqnarray}
when $\mu > m$. Finally, we point out that introducing a similar
asymmetric hopping $\delta\kappa$ in the case of uniform field
results in a Fermi surface in the weak-pairing regime.

\section{Gapless Surface Majorana fermions}

\begin{figure}
\includegraphics[width=0.492\columnwidth]{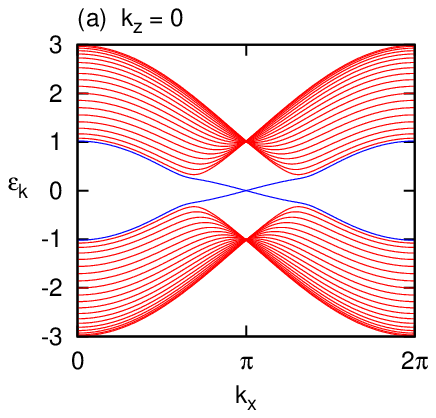}
\includegraphics[width=0.492\columnwidth]{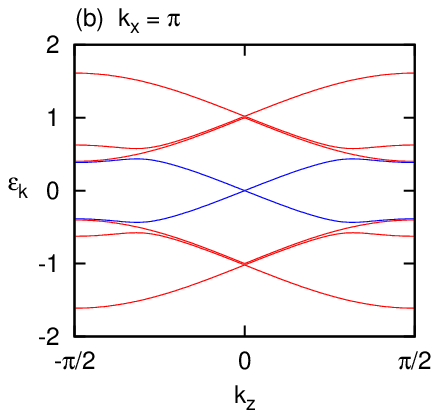}
\caption{\label{fig:surface} (Color online) Spectrum of a finite
system with $L = 10$ layers along $y$ direction. We consider the
symmetric case with $J_x = J_y = J_z \equiv J$. Coupling constant
along vertical links is set to $J_{\parallel} = 0.35 J$, and the
strength of the uniform field is $\kappa = 0.05 J$. Panels (a) and
(b) show dispersion along $k_x$ and $k_z$ directions, respectively,
of the 2D surface Brillouin zone.}
\end{figure}

Analogous to the gapless chiral edge modes in the non-Abelian phase
of 2D Kitaev model, \cite{kitaev06} the nontrivial topology of the
$\nu = \pm 1$ quantum ground state in our 3D model manifests itself
through the appearance of gapless Majorana fermions at the
sample surface. To study the properties of these surface modes, we
solve the Majorana hopping problem (\ref{eq:H-cc}) and
(\ref{eq:H-cc2}) in a finite geometry with periodic boundary
conditions along $x$ and $z$ directions and open boundary condition
along $y$ direction. We assume that the $x$-axis is parallel to the
zigzag direction of the honeycomb lattice. As Fig.~\ref{fig:surface}
shows, in addition to gapped states in the bulk, the spectrum of a
finite system contains additional surface modes crossing the bulk
gap.

The properties of these surface states can be understood
analytically using the edge modes of 2D Kitaev model, whose
existence has been demonstrated in Ref.~\onlinecite{kitaev06}. These
boundary modes are confined to the edge of the honeycomb lattice and
possess a definite chirality depending on the sign of magnetic field
$h$. Without loss of generality, we assume $\kappa\sim \lambda^2 h >
0$, hence $\epsilon_{\perp}(k_x) > 0$ for positive $k_x$ in the 1D
Brillouin zone. The Hamiltonian describing the edge states is given
by
\begin{eqnarray}
    \label{eq:H-edge}
    \mathcal{H}_{\rm edge} = \frac{1}{2}\sum_{k_x} \epsilon_{\perp}(k_x)
    \,\chi(-k_x)\, \chi(k_x),
\end{eqnarray}
where $\chi(k_x)$ and $\chi(-k_x) = \chi^\dagger(k_x)$ for $k_x>0$
are annihilation and creation operators, respectively, of the
Majorana edge modes. The spectrum of the edge states has the form
$\epsilon_{\perp}(k_x) \approx 12\kappa \sin k_x$ in the vicinity of
the 1D Fermi point $k^*_x = \pi$, and gradually merges with the bulk
spectrum as $k_x$ moves away from $k^*_x$. \cite{kitaev06}

It is interesting to note that the $J_{\parallel} = 0$ limit of the
3D model (\ref{eq:h}) can be viewed as a collection of decoupled
honeycomb layers. Due to the staggered magnetic field, the edge
states on even and odd numbered layers have opposite chirality $\nu
=\pm {\rm sgn}\,\kappa$; the corresponding quasiparticle operators
are denoted as $\chi_+(k_x,2n)$ and $\chi_-(k_x,2n+1)$,
respectively. The ensemble of the edge modes is described by
Hamiltonian:
\begin{eqnarray}
    \mathcal{H}_{\perp} = \frac{1}{2}\sum_n \sum_{k_x}
    \epsilon_{\perp}(k_x) \bigl[\chi_+(-k_x,2n)\chi_+(k_x,2n) \nonumber \\
     -\chi_-(-k_x,2n+1)\chi_-(k_x,2n+1)\bigr].
\end{eqnarray}
For odd-numbered layers, the quasiparticle annihilation operator is
given by $\chi_-(k_x, 2n+1)$ with negative $k_x$. A nonzero
$J_{\parallel}$ introduces coupling between adjacent layers:
\begin{eqnarray}
    \mathcal{H}_{\parallel} = iJ_{\parallel}\sum_{n}\sum_{k_x}
    \bigl[\chi_-(-k_x,2n+1)\chi_+(k_x,2n) \nonumber \\
    + \chi_+(-k_x,2n)\chi_-(k_x,2n-1)\bigr].
\end{eqnarray}
After Fourier transformation with respect to $z$ coordinate,
we obtain the Hamiltonian for surface modes
\begin{eqnarray}
    \mathcal{H}_{\rm surf} = \frac{1}{2}\sum_{k_x,k_z} \psi^\dagger_{k_x,
    k_z} \bigl[\epsilon_{\perp}(k_x)\sigma^z +
    \epsilon_{\parallel}(k_z) \sigma^x\bigr] \psi_{k_x,k_z},
\end{eqnarray}
where $\psi_{k_x,k_z} = \bigl(\chi_+(k_x,k_z),
\chi_-(k_x,k_z)\bigr)^T$, and the off-diagonal coupling
$\epsilon_{\parallel}(k_z) = 2J_{\parallel} \sin k_z$. The spectrum
can be easily obtained $\epsilon_{\rm surf} =
\bigl(\epsilon_{\parallel}^2 + \epsilon_{\perp}^2\bigr)^{1/2}$.
Consistent with the numerical calculation shown in
Fig.~\ref{fig:surface}, the surface spectrum has a conic singularity
at the 2D Fermi point $(k^*_x, k^*_z) = (\pi,0)$.

In the continuum approximation, the low-energy surface modes close
to the 2D Fermi point obeys a gapless Dirac-like Hamiltonian $H_{\rm
Dirac} = -i( u_{\perp} \sigma^z \partial_x + u_{\parallel} \sigma^x
\partial_z)$. The existence of gapless surface Majorana modes is a
direct consequence of the nontrivial ground-state topology when the
3D bulk sample is terminated by a 2D boundary.
\cite{callan85,fradkin86} Due to its topological nature, the surface
Dirac cone is stable against perturbations respecting the effective
TRS.

\section{Discussion}

To summarize, we have constructed an interacting bosonic model on a
three-dimensional layered honeycomb lattice which displays
topologically nontrivial ground states. Our approach is based on the
$\Gamma$-matrix generalization of Kitaev's original spin-1/2 model.
\cite{levin03,hamma05,yao09,wu09,ryu09,nussinov09} We show that the
ground state in the weak-pairing phase remains gapless in the
presence of a small magnetic field. The fermion spectrum of this
critical phase is characterized by four topologically protected
Fermi points. Upon increasing the field strength, Fermi points with
opposite winding numbers move toward each other and eventually
annihilate at a critical field, signaling the transition into a
topologically trivial phase with a gapped spectrum.

On the other hand, with the sign of magnetic field staggered along
successive layers, the bulk spectrum of the weak-pairing phase
immediately acquires a gap. An effective TRS is restored thanks to
the staggering of the magnetic field. The corresponding quantum
ground state is characterized by an integer winding invariant and
possesses nontrivial topological properties, a manifestation of
which is the appearance of protected gapless surface Majorana modes
when the sample is terminated by a two-dimensional surface.

It is worth noting that our model provides an alternative example in
which a 3D topological insulator emerges from an interacting bosonic
Hamiltonian. More importantly, the TRS essential to the existence of
3D topological insulators is realized in our model through a generic
physical mechanism, instead of relying on special symmetries of the
$\Gamma$-matrix representation in the recently proposed
diamond-lattice model. \cite{ryu09,wu09} We also remark that despite
the experimental difficulty of realizing the Kitaev model and its
variants, the study of these exactly solvable models has enriched
our understanding of the physics of topological phases. Finally, we
would like to point out that it remains unclear whether an emergent
topological insulator can be realized in a spin-1/2 Kitaev model on
a 3D trivalent lattice.
\\

\begin{acknowledgments}
The author acknowledges B. B\'eri, R. Moessner, N. Perkins, Tieyan
Si, Hong Yao, Sungkit Yip for helpful discussions and comments, and
also the visitors program at Max-Planck-Institut f\"ur Physik
komplexer Systeme, Dresden, Germany. In particular, I would like to
thank Benjamin B\'eri for pointing out the possibility of
topologically stable Fermi lines in the DIII Hamiltonian studied in
this paper.
\end{acknowledgments}

\end{document}